\documentclass[12pt, preprint]{aastex}

\usepackage{graphicx}
\usepackage{natbib}
\usepackage{lscape}
\bibpunct{(}{)}{;}{a}{}{,}
\shorttitle{Neutron-Capture Elements in Magellanic Cloud PNe}
\shortauthors{Sterling et al.}

\begin{document}

\title{Neutron-Capture Element Abundances in Magellanic Cloud Planetary Nebulae\footnote{This paper includes data obtained with the 6.5-m Magellan Telescopes located at Las~Campanas Observatory, Chile, and with the Gemini-South Telescope at Cerro Pachon, Chile.}}

\author{A.\ L.\ Mashburn\altaffilmark{1}, N.\ C.\ Sterling\altaffilmark{1}, S.\ Madonna\altaffilmark{2}, Harriet L.\ Dinerstein\altaffilmark{3}, I.\ U.\ Roederer\altaffilmark{4, 5}, T.\ R.\ Geballe\altaffilmark{6}}

\altaffiltext{1}{Department of Physics, University of West Georgia, 1601 Maple Street, Carrollton, GA 30118, USA; awhite15@my.westga.edu, nsterlin@westga.edu}
\altaffiltext{2}{Instituto de Astrof\'{i}sica de Canarias, E-38205 La Laguna, Tenerife, Spain; Universidad de La Laguna, Dpto. Astrof\'isica, E-38206 La Laguna, Tenerife, Spain; smadonna@iac.es}
\altaffiltext{3}{Department of Astronomy, University of Texas, 2515 Speedway, C1400, Austin, TX 78712-1205, USA; harriet@astro.as.utexas.edu}
\altaffiltext{4}{Department of Astronomy, University of Michigan, 1085 South University Avenue, Ann Arbor, MI 48109, USA; iur@umich.edu}
\altaffiltext{5}{Joint Institute for Nuclear Astrophysics and Center for the Evolution of the Elements (JINA-CEE), USA}
\altaffiltext{6}{Gemini Observatory, 670 N.\ A'ohoku Place, Hilo, HI 96720, USA; tgeballe@gemini.edu}

\begin{abstract}

We present near-infrared spectra of ten planetary nebulae (PNe) in the Large and Small Magellanic Clouds (LMC and SMC), acquired with the FIRE and GNIRS spectrometers on the 6.5-m Baade and 8.1-m Gemini South Telescopes, respectively.  We detect Se and/or Kr emission lines in eight of these objects, the first detections of \emph{n}-capture elements in Magellanic Cloud PNe.  Our abundance analysis shows large \emph{s}-process enrichments of Kr (0.6--1.3~dex) in the six PNe in which it was detected, and Se is enriched by 0.5--0.9~dex in five objects.  We also estimate upper limits to Rb and Cd abundances in these objects.  Our abundance results for the LMC are consistent with the hypothesis that PNe with 2--3~M$_{\odot}$ progenitors dominate the bright end of the PN luminosity function in young gas-rich galaxies.  We find no significant correlations between \emph{s}-process enrichments and other elemental abundances, central star temperature, or progenitor mass, though this is likely due to our small sample size.  We determine S abundances from our spectra and find that $[$S/H$]$ agrees with $[$Ar/H$]$ to within 0.2~dex for most objects, but is lower than $[$O/H$]$ by 0.2--0.4~dex in some PNe, possibly due to O enrichment via third dredge-up.  Our results demonstrate that \emph{n}-capture elements can be detected in PNe belonging to nearby galaxies with ground-based telescopes, allowing \emph{s}-process enrichments to be studied in PN populations with well-determined distances.

\end{abstract}


\section{INTRODUCTION} \label{intro}

Emission lines of neutron(\emph{n})-capture elements (atomic number $Z>30$) were first identified in a planetary nebula (PN) in 1994 \citep{pb94}, and since have been detected in more than 100 Galactic PNe \citep[e.g.,][]{sharpee07, sterling08, garcia-rojas15}.  Trans-iron elements can be produced by slow \emph{n}-capture nucleosynthesis (the ``\emph{s}-process'') in asymptotic giant branch (AGB) stars, and transported to the stellar envelope by third dredge-up (TDU) before being expelled via stellar winds and PN ejection \citep{karakas14}.  Comparisons of empirically-determined \emph{s}-process enrichments in PNe to theoretical predictions provide valuable constraints to models of AGB nucleosynthesis \citep{karakas09, sterling16}.

To date, nebular \emph{n}-capture element abundance determinations have almost exclusively been limited to Galactic PNe, whose primarily statistical distances can have substantial uncertainties \citep[though improved calibrations to statistical distance scales show promise for better accuracies;][and references therein]{frew16}.  Because of the uncertain distances, it has not been possible to study \emph{s}-process enrichments along the PN luminosity function \citep[PNLF,][]{jacoby89} -- which prevents robust estimates of the fraction of PNe that are \emph{s}-process enriched \citep{sterling08} -- or as a function of initial stellar mass.

Extragalactic PNe do not suffer from the distance uncertainties that plague Galactic objects.  However this advantage comes at a cost, as the large distances of these PNe render the detection of faint emission lines difficult.  Nevertheless, it is possible to detect \emph{n}-capture elements in Local Group PNe with sufficiently large-aperture telescopes.  

The Large and Small Magellanic Clouds (LMC and SMC) are optimal targets for such a study, given their relative proximity \citep[50 and 60~kpc, respectively;][]{keller06}, minimal foreground extinction, and relatively well-studied PN populations.  A significant fraction of LMC and SMC PNe have been identified \citep{reid14, dravskovic15}, and elemental abundances have been determined in a large number of these objects \citep[e.g.,][and references therein]{ld06}.  In addition, progenitor star masses have been estimated for some PNe in these galaxies \citep{villaver03, villaver04, villaver07}.

In this letter we present the detection of near-infrared $[$\ion{Kr}{3}$]$ and $[$\ion{Se}{4}$]$ emission lines in ten bright LMC and SMC PNe.  To our knowledge these are the first detections of \emph{n}-capture elements in extragalactic PNe other than the Sagittarius Dwarf \citep{wood06, otsuka11}.

\section{OBSERVATIONS AND REDUCTIONS}

\defcitealias{shaw10}{S10}
\defcitealias{tsamis03}{T03}
\defcitealias{ld06}{LD06}
\defcitealias{meatheringham91b}{MD91}

In Table~\ref{objects} we provide an observing log and nebular and stellar parameters for our sample.  Nine of the ten PNe were observed with the Folded-Port InfraRed Echellette (FIRE) spectrograph \citep{simcoe13} on the 6.5-m Baade Telescope at Las Campanas Observatory.  We used a 0\farcs75 slit width to provide a resolution $R=4800$ in echelle mode, covering the spectral range 0.83--2.45~$\mu$m.  Because the targets have diameters comparable to or smaller than the slit width, light loss primarily occurred due to seeing conditions, which were typically less than 1\arcsec\ but ranged from 2\arcsec --4\arcsec\ for LMC~SMP~73 and SMC~SMP~15.  We nodded along the slit in ABBA sequences for maximum observing efficiency.  The data were reduced using the FIREHOSE IDL reduction pipeline\footnote{Available at http://web.mit.edu/\~{}rsimcoe/www/FIRE/}.  Th-Ar lamps were used to wavelength calibrate the spectra, and A0V standard stars were observed for each object to perform relative flux calibrations and telluric corrections.

\begin{deluxetable}{llccccccccrcc}
\tablecolumns{13}
\tabletypesize{\scriptsize}
\tablewidth{0pc} 
\tablecaption{Observing Log and Nebular Properties}
\tablehead{
\colhead{} & \colhead{PN} & \colhead{Date} & \colhead{} & \colhead{Int.} & \colhead{$T_{\mathrm{e}}[$\ion{O}{3}$]$} & \colhead{$n_e$} & \colhead{} & \colhead{} & \colhead{Log} & \colhead{Log} & \colhead{$T_{\rm eff}$\tablenotemark{c}} & \colhead{$M_{\rm init}$\tablenotemark{c}} \\
\colhead{Galaxy} & \colhead{Name} & \colhead{Observed} & \colhead{Inst.} & \colhead{Time (s)} & \colhead{(10$^3$ K)} & \colhead{(10$^3$ cm$^{-3}$)} & \colhead{$c_{\rm H\beta }$} & \colhead{$m_{5007}$\tablenotemark{a}} & \colhead{C/O\tablenotemark{b}} & \colhead{N/O} & \colhead{(10$^3$ K)} & \colhead{(M$_{\odot}$)}}
\startdata
LMC & SMP 6  & 2013-01-22 & FIRE  & 4600 & 13.3 & 11.8 & 0.04 & 15.53  & \nodata & --0.99 & 140.0 & \nodata \\
    & SMP 47 & 2013-01-22 & FIRE  & 2100 & 14.7 & 4.8  & 0.42 & 15.28  & 0.37    & 0.45   & 150.0 & \nodata \\
    & SMP 62 & 2006-08-16 & GNIRS & 1120 & 15.9 & 3.4  & 0.06 & 14.67  & --0.85  & --0.26 & 100.0 & \nodata \\
    & SMP 63 & 2013-01-21 & FIRE  & 3840 & 11.9 & 7.4  & 0.11 & 15.17  & 0.01    & --0.34 & 38.8  & 1.5--2.0 \\
    & SMP 73 & 2013-08-12 & FIRE  & 5600 & 11.7 & 4.5  & 0.34 & 14.93  & 0.18    & --0.63 & 135.0 & \nodata \\
    & SMP 85 & 2013-01-21 & FIRE  & 2400 & 11.7 & 31.4 & 0.42 & 16.15  & 0.64    & --0.74 & 46.0  & \nodata  \\
    & SMP 99 & 2013-01-21 & FIRE  & 4320 & 12.7 & 2.29 & 0.35 & 14.98  & 0.28    & --0.60\tablenotemark{d} & 124.0 & \nodata \\
SMC & SMP 15 & 2013-08-11 & FIRE  & 5400 & 12.0 & 5.0  & 0.04 & 15.67  & 0.12    & --0.37 & 58.0  & \nodata  \\
    & SMP 17 & 2013-08-12 & FIRE  & 8000 & 12.2 & 2.9  & 0.06 & 15.52  & 0.19    & --0.83 & 58.4  & 1.0      \\
    & SMP 20 & 2013-08-11 & FIRE  & 7200 & 13.8 & 3.9  & 0.0  & 16.12  & 0.51    & --0.79 & 86.5  & 1.0--1.5 \\
    & SMP 20 & 2006-08-16 & GNIRS & 1120 &      &      &      &        &         &        &       &          \\
\tableline
\enddata
\label{objects}
\tablecomments{Nebular temperatures, densities, extinction coefficients, N/O, and (unless specified) C/O abundances are from \citet[][hereafter S10]{shaw10} for SMP~17 and 20, \citet[][T03]{tsamis03} for SMP~63, \citet[][MD91]{meatheringham91b} for SMP~99, and \citet[][LD06]{ld06} for the remaining PNe.}
\tablenotetext{b}{Apparent $[$\ion{O}{3}$]$~5007 magnitudes computed from absolute fluxes (corrected for foreground extinction) measured with the \emph{Hubble Space Telescope} \citep{stanghellini03, shaw06}, with the exceptions of SMP~85 and SMP~99 \citepalias{ld06}, using the relation $m_{5007} = -2.5\mathrm{log}F_{5007} - 13.74$ \citep{jacoby89}.}
\tablenotetext{b}{C/O abundances are from the references above, with the exceptions of SMP~62 and SMP~17 \citep{aller87}, SMP~85 \citep{dopita94}, and SMP~15 and SMP~20 \citep{stanghellini09}.}
\tablenotetext{c}{References for central star temperatures: SMP~62 \citep{aller87}, SMP~85 \citep{dopita94}, SMP~63 \citep{villaver03}, SMP~17 and 20 \citep{villaver04}, and \citet{dopita91b} for all other PNe.  Estimated progenitor masses $M_{\rm init}$ are from \citet{villaver03, villaver04}.}
\tablenotetext{d}{N/O ratio from \citet{dopita91b}}
\end{deluxetable}

LMC~SMP~62 and SMC~SMP~20 were observed in the $K$~band with the Gemini Near-InfraRed Spectrograph (GNIRS) on the 8.1-m Gemini South telescope.  We used the 111~l/mm grating in third order with a 0\farcs45$\times$99\arcsec\ slit, for an effective resolving power of $R=4000$ in the wavelength range 2.1--2.3~$\mu$m.  The data were taken in queue mode under observing program GS-2006B-Q-51.  We beam-switched by nodding the target along the slit.  The wavelength scale was established with an Ar arc lamp, and A0V standard stars were observed for flux calibration and telluric absorption corrections.  The data were reduced using the FIGARO software package \citep{shortridge93}.  Fluxes and upper limits in the Gemini $K$~band spectrum of SMP~20 agree well with our FIRE data, and we restrict our analysis to the FIRE data for this PN.

\section{LINE MEASUREMENTS AND ABUNDANCE ANALYSIS} \label{abundances}

\begin{figure}[ht!]
\epsscale{0.8}
\plotone{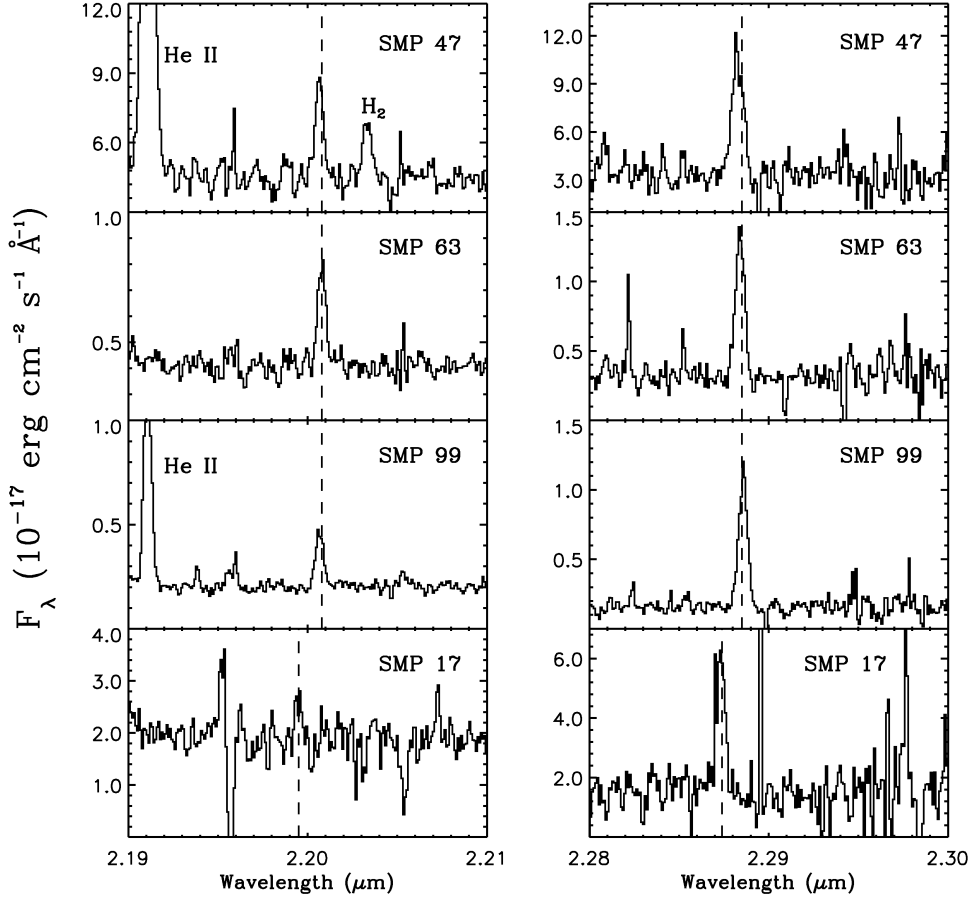}
\caption{$[$\ion{Kr}{3}$]$~2.1986 (left panels) and $[$\ion{Se}{4}$]$~2.2864~$\mu$m (right panels) detections in selected LMC PNe and SMC~SMP~17.  Vertical dashed lines correspond to the typical observed wavelength of each feature in LMC/SMC PNe.}
\label{profiles}
\end{figure}

The FIRE spectra are very rich, with 80--110 emission lines detected in LMC objects and 60--85 in SMC PNe.  We detect metal lines including $[$\ion{C}{1}$]$, $[$\ion{P}{2}$]$, $[$\ion{S}{2}$]$, $[$\ion{S}{3}$]$, $[$\ion{Fe}{3}$]$, $[$\ion{Kr}{3}$]$, and $[$\ion{Se}{4}$]$, and several LMC PNe exhibit H$_2$ lines.  We have also detected $[$\ion{Kr}{6}$]$~1.2333~$\mu$m\footnote{All wavelengths reported in this paper are vacuum wavelengths.} in LMC~SMP~47 and SMP~99, and will discuss this identification in a forthcoming paper.  Notably, Kr and/or Se were detected in all seven of the observed LMC PNe, and in one of the three SMC objects (Figure~\ref{profiles}).

We measured line fluxes by integrating under the profile of each line above a local continuum using IDL routines, varying the continuum placement to estimate flux uncertainties.  Gaussian fits were performed for blended features.  We also measured 3-$\sigma$ upper limits to the $[$\ion{Rb}{4}$]$~1.5973, $[$\ion{Cd}{4}$]$~1.7203, and $[$\ion{Ge}{6}$]$~2.1930~$\mu$m lines identified by \citet{sterling16} and to $[$\ion{Se}{3}$]$ features.  Line fluxes and intensities relative to \ion{H}{1}~Br$\gamma$ are reported in Table~\ref{linelist}.

\begin{deluxetable}{lccccl}
\tablecolumns{6}
\tablewidth{0pc} 
\tablecaption{Line Identifications and Intensities}
\tablehead{
\colhead{} & \colhead{Observed} & \colhead{Lab.} & \colhead{$F$/$F$(Br$\gamma$)} & \colhead{$I$/$I$(Br$\gamma$)} & \colhead{}\\
\colhead{Line ID} & \colhead{$\lambda$ ($\mu$m)} & \colhead{$\lambda$ ($\mu$m)} & \colhead{($\times 100$)} & \colhead{($\times 100$)} & \colhead{Comments}}
\startdata
\multicolumn{6}{c}{\textbf{LMC SMP 6}} \\
\cline{1-6}
\ion{H}{1}     &     0.8368 & 0.8361 &   (3.66$\pm$0.93)E+00 &   (6.59$\pm$1.68)E+00 & : \\
\ion{H}{1}     &     0.8401 & 0.8395 &   (4.81$\pm$1.09)E+00 &   (8.61$\pm$1.94)E+00 & \\
\ion{H}{1}     &     0.8447 & 0.8440 &   (8.56$\pm$1.33)E+00 &   (1.52$\pm$0.24)E+01 & \\
\ion{H}{1}     &     0.8477 & 0.8470 &   (9.20$\pm$1.47)E+00 &   (1.63$\pm$0.26)E+01 & \\
\ion{H}{1}     &     0.8513 & 0.8505 &   (1.07$\pm$0.20)E+01 &   (1.89$\pm$0.36)E+01 & \\
\ion{H}{1}     &     0.8555 & 0.8548 &   (7.64$\pm$1.70)E+00 &   (1.33$\pm$0.30)E+01 & \\
\ion{H}{1}     &     0.8608 & 0.8601 &   (1.16$\pm$0.11)E+01 &   (2.00$\pm$0.20)E+01 & \\
\vdots      &  \vdots & \vdots &     \vdots              &         \vdots           & \vdots   \\
\cline{1-6}
\tableline
\enddata
\label{linelist}
\tablecomments{Table~\ref{linelist} is published in its entirety as an online supplement.  A portion is shown here for guidance regarding its form and content.  Measured fluxes and intensities are on the scale $F$(\ion{H}{1}~Br~$\gamma$)~=~$I$(\ion{H}{1}~Br~$\gamma$)~=~100.  Marginal detections are marked with a colon.}
\end{deluxetable}

For our abundance analysis, we adopt extinction coefficients, temperatures, densities, and ionic abundances used in ionization correction factors (ICFs) from the literature (see Table~\ref{objects} for references).  We adopt 10\% uncertainties for He ionic abundances from the literature, and 30\% for those of O and Ar.  \citetalias{meatheringham91b} and \citetalias{ld06} do not report ionic abundances, and we derived these values from their listed intensities.

We calculated ionic and elemental abundances with the PyNeb analysis package \citep{luridiana15}, using the atomic data sources of \citet[][see their Table~5]{garcia-rojas15}, with the exceptions of Rb$^{3+}$, Cd$^{3+}$, and Ge$^{5+}$ for which we use the atomic data of \citet{sterling16}.  To minimize uncertainties due to errors in the flux calibration or adopted extinction coefficients, we computed ionic abundances relative to nearby \ion{H}{1} lines: 11--3 for lines with wavelengths $\leq$1.0~$\mu$m, Pa$\beta$ for the $J$~band, Br$\zeta$ for $H$~band lines, and Br$\gamma$ in the $K$~band.

The $[$\ion{Se}{4}$]$~2.2864~$\mu$m line can be contaminated by H$_2$~3-2~S(2)~2.2870~$\mu$m in PNe with fluorescent H$_2$ emission \citep{dinerstein01}, and this must be accounted for when computing Se$^{3+}$/H$^+$ abundances.  The strengths of H$_2$ lines from the $v=2$ and $v=3$ levels in LMC~SMP~47 and 85 are indicative of fluorescent excitation of moderately dense gas. For such conditions, the intensity of H$_2$~3-2~S(2) is about 0.8 times that of the 3-2~S(3) line \citep[e.g., model 14 of][for which $n = 3 \times 10^3$~cm$^{-3}$]{bvd87}.  We use these values to correct the measured fluxes at 2.287~$\mu$m for the contribution of H$_2$. The absence of the 3-2~S(3) line indicates that such corrections are not needed for the other PNe.

We report ionic abundances in Table~\ref{ionicf}.  The error bars include uncertainties in the line fluxes, and assumed error bars of 1000~K in $T_{\rm e}$ and 20\% for $n_{\rm e}$ values from the literature.  All abundance uncertainties were propagated via Monte Carlo simulations.

\begin{landscape}
\begin{deluxetable}{lccccccccccc}
\tablecolumns{11}
\tablewidth{0pc} 
\tabletypesize{\scriptsize}
\tablecaption{Ionic Abundances and Ionization Correction Factors} 
\tablehead{
\colhead{} & \colhead{LMC} & \colhead{LMC} & \colhead{LMC} & \colhead{LMC} & \colhead{LMC} & \colhead{LMC} & \colhead{LMC} & \colhead{SMC} & \colhead{SMC} & \colhead{SMC} \\
\colhead{} & \colhead{SMP 6} & \colhead{SMP 47} & \colhead{SMP 62 (GNIRS)} & \colhead{SMP 63} & \colhead{SMP 73} & \colhead{SMP 85} & \colhead{SMP 99} & \colhead{SMP 15} & \colhead{SMP 17} & \colhead{SMP 20}}
\startdata
\multicolumn{11}{c}{\textit{Derived Ionic Abundances}} \\
\cline{1-11}
S$^{+}$/H$^+$   & (1.9$\pm$1.0)E-07 & (3.6$\pm$1.5)E-07 & (3.6$\pm$1.1)E-07\tablenotemark{a} & (1.2$\pm$0.9)E-07 & (4.9$\pm$3.3)E-07 & (1.8$\pm$1.3)E-07 & (2.1$\pm$1.3)E-07 & (2.5$\pm$2.0)E-07 & (1.2$\pm$0.6)E-07 & (2.7$\pm$1.6)E-08 \\
S$^{2+}$/H$^+$  & (1.2$\pm$0.2)E-06 & (2.7$\pm$0.4)E-06 & (8.1$\pm$2.4)E-07\tablenotemark{a} & (1.7$\pm$0.4)E-06 & (2.0$\pm$0.5)E-06 & (1.4$\pm$0.3)E-06 & (1.5$\pm$0.3)E-06 & (1.4$\pm$0.3)E-06 & (1.0$\pm$0.3)E-06 & (2.5$\pm$0.4)E-07 \\
Ge$^{5+}$/H$^+$ & $\leq$6.6E-11      & $\leq$3.3E-10      & $\leq$2.4E-10                       & $\leq$1.4E-10      & $\leq$2.2E-10      & $\leq$1.9E-10      & $\leq$1.5E-10      & $\leq$2.2E-10      & $\leq$5.5E-11      & $\leq$1.5E-10      \\
Se$^{2+}$/H$^+$ & $\leq$1.0E-09      & $\leq$1.4E-09      & \nodata                              & $\leq$4.7E-09      & $\leq$9.1E-09      & $\leq$6.6E-09      & $\leq$2.9E-09      & $\leq$7.9E-09      & $\leq$1.2E-09      & $\leq$1.2E-09      \\
Se$^{3+}$/H$^+$ & (7.2$\pm$1.0)E-10 & (1.9$\pm$0.6)E-10 & (7.4$\pm$2.6)E-11                  & (7.2$\pm$0.8)E-10 & (1.9$\pm$0.3)E-09 & (3.1$\pm$0.9)E-10 & (2.0$\pm$0.2)E-09 & $\leq$3.4E-11      & (7.5$\pm$1.9)E-10 & $\leq$2.4E-11      \\
Kr$^{2+}$/H$^+$ & (7.7$\pm$1.6)E-10 & (1.5$\pm$0.3)E-10 & $\leq$3.6E-10                       & (8.7$\pm$1.4)E-10 & $\leq$5.8E-10      & (2.6$\pm$0.7)E-09 & (2.2$\pm$0.2)E-09 & $\leq$3.1E-10      & (3.7$\pm$1.3)E-10 & $\leq$2.4E-10      \\
Rb$^{3+}$/H$^+$ & $\leq$1.6E-10      & $\leq$2.3E-10      & \nodata                              & $\leq$3.7E-10      & $\leq$8.7E-09      & $\leq$4.1E-10      & $\leq$3.2E-10      & $\leq$9.8E-10      & $\leq$8.8E-11      & $\leq$1.9E-10      \\
Cd$^{3+}$/H$^+$ & $\leq$5.7E-11      & $\leq$1.2E-10      & \nodata                              & $\leq$1.4E-10      & $\leq$7.6E-10      & $\leq$2.3E-10      & $\leq$1.6E-10      & $\leq$4.6E-10      & $\leq$1.2E-10      & $\leq$7.7E-11      \\
\cutinhead{\textit{Ionic Abundances from the Literature and ICFs}\tablenotemark{b}} \\
He$^+$/H$^+$   & (6.0$\pm$0.6)E-02 & (8.2$\pm$0.8)E-02 & (7.3$\pm$0.7)E-02  & (1.1$\pm$0.1)E-01 & (7.0$\pm$0.7)E-02 & (7.6$\pm$0.8)E-02 & (8.8$\pm$0.9)E-02 & (9.3$\pm$0.9)E-02 & (1.4$\pm$0.1)E-01 & (1.4$\pm$0.1)E-01 \\
He$^{2+}$/H$^+$ & (3.9$\pm$0.4)E-02 & (3.8$\pm$0.4)E-02 & (2.5$\pm$0.3)E-02 & (3.0$\pm$0.3)E-04 & (2.2$\pm$0.2)E-02 & \nodata             & (2.1$\pm$0.2)E-02 & (2.3$\pm$0.2)E-02 & (1.0$\pm$0.1)E-03 & \nodata             \\
O$^+$/H$^+$    & (4.9$\pm$1.5)E-05 & (9.7$\pm$2.9)E-06 & (8.5$\pm$2.5)E-06 & (4.5$\pm$1.3)E-06 & (3.1$\pm$0.9)E-05 & (5.0$\pm$1.5)E-05 & (3.1$\pm$0.9)E-05 & (6.7$\pm$2.0)E-06 & (6.8$\pm$2.0)E-06 & (2.0$\pm$0.6)E-06 \\
O$^{2+}$/H$^+$  & (1.6$\pm$0.5)E-04 & (1.1$\pm$0.3)E-04 & (1.0$\pm$0.3)E-04 & (1.9$\pm$0.6)E-04 & (3.1$\pm$0.9)E-04 & (7.6$\pm$2.3)E-05 & (2.0$\pm$0.6)E-04 & (1.1$\pm$0.3)E-04 & (1.6$\pm$0.5)E-04 & (5.4$\pm$1.6)E-05 \\
Ar$^{2+}$/H$^+$ & (3.4$\pm$1.0)E-07 & (6.6$\pm$2.0)E-07 & \nodata             & (6.4$\pm$1.9)E-07 & (6.4$\pm$1.9)E-07 & (3.4$\pm$1.0)E-07 & (5.8$\pm$1.7)E-07 & (2.9$\pm$0.9)E-07 & (2.6$\pm$0.8)E-07 & (1.0$\pm$0.3)E-07 \\
ICF(O)         & 1.38$\pm$0.27     & 1.27$\pm$0.21     & 1.19$\pm$0.17       & 1.00               & 1.18$\pm$0.16     & 1.00               & 1.13$\pm$0.13     & 1.14$\pm$0.14      & 1.00               & 1.00 \\
ICF(Ar)        & 1.68$\pm$1.09     & 1.89$\pm$1.22     & \nodata             & 1.68$\pm$1.09     & 1.72$\pm$1.12     & 1.08$\pm$0.71     & 1.54$\pm$1.00     & 1.77$\pm$1.15      & 1.61$\pm$1.05     & 1.62$\pm$1.06 \\
ICF(S)         & 1.55$\pm$0.27       & 2.06$\pm$0.33       & 1.97$\pm$0.31       & 2.24$\pm$0.36       & 1.84$\pm$0.30       & 1.00                & 1.54$\pm$0.25       & 2.06$\pm$0.33       & 1.95$\pm$0.31  & 1.00                       \\
ICF(Se)        & 2.98$\pm$2.15       & 2.09$\pm$1.72       & 1.91$\pm$1.55       & 1.37$\pm$1.48       & 1.92$\pm$1.74       & 2.67$\pm$1.78       & 1.95$\pm$1.78       & 1.73$\pm$1.44       & 1.42$\pm$1.63  & 1.40$\pm$1.18              \\
ICF(Kr)        & 2.83$\pm$1.74       & 3.73$\pm$2.01       & 4.58$\pm$3.70       & 3.85$\pm$2.61       & 3.66$\pm$2.51       & 1.64$\pm$0.58       & 2.74$\pm$1.65       & 3.88$\pm$2.38       & 3.46$\pm$2.39  & 1.59$\pm$0.47              \\
ICF(Rb, Cd)    & 1.80$\pm$0.78       & 1.38$\pm$0.61       & \nodata             & 1.02$\pm$0.43       & 1.30$\pm$0.56       & 1.65$\pm$0.61       & 1.31$\pm$0.54       & 1.20$\pm$0.52       & 1.05$\pm$0.44  & 1.04$\pm$0.43              \\
\tableline
\enddata
\label{ionicf}
\tablenotetext{a}{Derived from \citetalias{ld06} intensities}
\tablenotetext{b}{See text}
\end{deluxetable}
\end{landscape}

To convert ionic abundances to elemental abundances, we employ the ICF formulae of \citet{delgado-inglada14} for light elements and those of \citet{sterling15, sterling16} for \emph{n}-capture elements.  Uncertainties to the ICFs are the recommendations of \cite{delgado-inglada14} for O, S, and Ar, and were propagated from the ionic and elemental abundances used in the ICF prescriptions for \emph{n}-capture elements.  In the case of the Kr ICF, we use Equation~1 of \citet{sterling15}, which depends on S$^{2+}$/S, rather than Equation~2 (which uses Ar$^{2+}$/Ar), since $[$\ion{Ar}{3}$]$ lines were not detected in SMP~62 \citepalias{ld06} and the derived Ar abundance in SMP~85 is larger than the solar value \citep{asplund09} and appears to be inaccurate.  The two equations produce Kr ICFs that agree to within 25\% for our targets except for SMC~SMP~20 (in which Kr is not detected), with no systematic trends.  Therefore this choice does not affect our results.

In Table~\ref{abund} we give elemental abundances relative to the solar values of \citet{asplund09}, with uncertainties accounting for those in the ionic abundances and ICFs.  Our derived O abundances agree with literature values \citepalias{meatheringham91b, tsamis03, ld06, shaw10} to within 25\% or better, while those for Ar show more scatter but agree to within 40\% for most PNe.  Our S abundances are factors of 2--14 lower than the values of \citetalias{ld06}, in line with the findings of \citet{bernard-salas08} and \citet{shaw10}.

\begin{landscape}
\begin{deluxetable}{lccccccccccc}
\tablecolumns{11}
\tablewidth{0pc} 
\tabletypesize{\scriptsize}
\tablecaption{Elemental Abundances} 
\tablehead{
\colhead{} & \colhead{LMC} & \colhead{LMC} & \colhead{LMC} & \colhead{LMC} & \colhead{LMC} & \colhead{LMC} & \colhead{LMC} & \colhead{SMC} & \colhead{SMC} & \colhead{SMC} \\
\colhead{} & \colhead{SMP 6} & \colhead{SMP 47} & \colhead{SMP 62} & \colhead{SMP 63} & \colhead{SMP 73} & \colhead{SMP 85} & \colhead{SMP 99} & \colhead{SMP 15} & \colhead{SMP 17} & \colhead{SMP 20}}
\startdata
$[$O/H$]$      & --0.23$\pm$0.12     & --0.50$\pm$0.12     & --0.57$\pm$0.12                      & --0.39$\pm$0.11     & --0.09$\pm$0.12     & --0.59$\pm$0.08     & --0.28$\pm$0.11     & --0.55$\pm$0.12     & --0.48$\pm$0.11     & --0.95$\pm$0.11 \\
$[$S/H$]$      & --0.80$\pm$0.17     & --0.32$\pm$0.17     & --0.76$\pm$0.18                      & --0.52$\pm$0.18     & --0.46$\pm$0.18     & --0.92$\pm$0.08     & --0.70$\pm$0.19     & --0.59$\pm$0.18     & --0.77$\pm$0.18     & --1.68$\pm$0.06 \\
$[$Ar/H$]$     & --0.65$\pm$0.26     & --0.30$\pm$0.26     & \nodata                              & --0.37$\pm$0.26     & --0.36$\pm$0.26     &   0.17$\pm$0.25     & --0.45$\pm$0.26     & --0.69$\pm$0.26     & --0.78$\pm$0.26     & --1.18$\pm$0.26 \\
\cline{1-11}
$[$Se/H$]$     & --0.01$\pm$0.25     & --0.74$\pm$0.30     & --1.19$\pm$0.30                      & --0.35$\pm$0.27     &   0.21$\pm$0.29     & --0.43$\pm$0.24     &   0.25$\pm$0.24     & $\leq$--1.57        & --0.32$\pm$0.32     & $\leq$--1.82   \\
$[$Se/S$]$     & 0.80$\pm$0.28       & --0.42$\pm$0.33     & --0.43$\pm$0.33                      &   0.17$\pm$0.30     &   0.67$\pm$0.32     & 0.50$\pm$0.25       & 0.95$\pm$0.28       & $\leq$--0.97         & 0.45$\pm$0.34      & $\leq$--0.14    \\
$[$Se/Ar$]$    & 0.64$\pm$0.33       & --0.44$\pm$0.36     & \nodata                              &   0.02$\pm$0.34     &   0.57$\pm$0.35     & --0.59$\pm$0.32     & 0.70$\pm$0.32       & $\leq$--0.87         & 0.47$\pm$0.37      & $\leq$--0.64    \\
$[$Kr/H$]$     & 0.09$\pm$0.22       & 0.49$\pm$0.20       & $\leq$--0.04                         & 0.27$\pm$0.23       & $\leq$0.08          & 0.38$\pm$0.16       & 0.52$\pm$0.21       & $\leq$--0.17        & --0.15$\pm$0.25       & $\leq$--0.67   \\
$[$Kr/S$]$     & 0.89$\pm$0.26       & 0.81$\pm$0.24       & $\leq$0.72                           & 0.79$\pm$0.27       & $\leq$0.53          & 1.30$\pm$0.17       & 1.23$\pm$0.26       & $\leq$0.42          & 0.62$\pm$0.29         & $\leq$1.01     \\
$[$Kr/Ar$]$    & 0.74$\pm$0.31       & 0.80$\pm$0.30       & \nodata                              & 0.64$\pm$0.32       & $\leq$0.44          & 0.21$\pm$0.28       & 0.97$\pm$0.31       & $\leq$0.52          & 0.63$\pm$0.33         & $\leq$0.51     \\
$[$Rb/H$]$     & $\leq$--0.07        & $\leq$--0.01        & \nodata                              & $\leq$0.06          & $\leq$0.53          & $\leq$0.31          & $\leq$0.10          & $\leq$0.55          & $\leq$--0.56          & $\leq$--0.23   \\
$[$Rb/S$]$     & $\leq$0.73          & $\leq$0.30          & \nodata                              & $\leq$0.58          & $\leq$0.99          & $\leq$1.23          & $\leq$0.80          & $\leq$1.14          & $\leq$0.21            & $\leq$1.45  \\
$[$Rb/Ar$]$    & $\leq$0.58          & $\leq$0.29          & \nodata                              & $\leq$0.43          & $\leq$0.89          & $\leq$0.14          & $\leq$0.54          & $\leq$1.25          & $\leq$0.22            & $\leq$0.95  \\
$[$Cd/H$]$     & $\leq$0.30          & $\leq$0.49          & \nodata                              & $\leq$0.44          & $\leq$1.28          & $\leq$0.87          & $\leq$0.60          & $\leq$1.03          & $\leq$0.38            & $\leq$0.19   \\
$[$Cd/S$]$     & $\leq$1.11          & $\leq$0.80          & \nodata                              & $\leq$0.96          & $\leq$1.74          & $\leq$1.79          & $\leq$1.31          & $\leq$1.62          & $\leq$1.15            & $\leq$1.87   \\
$[$Cd/Ar$]$    & $\leq$0.95          & $\leq$0.79          & \nodata                              & $\leq$0.81          & $\leq$1.64          & $\leq$0.70          & $\leq$1.05          & $\leq$1.73          & $\leq$1.16            & $\leq$1.37   \\
\tableline
\enddata
\label{abund}
\tablecomments{Abundances $[$X/H$]$~=~log(X/H)$_{\rm PN} -$~log(X/H)$_{\odot}$, computed from the ionic abundances and ICFs in Table~\ref{ionicf}.}
\end{deluxetable}
\end{landscape}

\section{ABUNDANCE PATTERNS AND ENRICHMENTS}

\subsection{Evidence for Third Dredge-Up and Choice of a Metallicity Reference} \label{lightelm}

TDU conveys C-rich and \emph{s}-process enriched material to the envelopes of AGB stars \citep{karakas14}, and therefore C-rich PNe can be expected to exhibit \emph{s}-process enrichments.  With the exceptions of LMC~SMP~62 \citep{aller87} and SMP~6 (no C abundance available), all PNe in our sample have C/O ratios of unity or larger, and thus experienced TDU.

To determine whether a PN is \emph{s}-process enriched it is necessary to compare \emph{n}-capture element abundances to that of an element representative of the metallicity $[$Fe/H$]$, since Fe abundances cannot be accurately determined in nebulae due to depletion into dust \citep[e.g.,][]{delgado-inglada14a}.  In LMC field giants, $\alpha$-elements such as O, Mg, Si, Ca, and Ti are approximately solar relative to Fe ($[\alpha$/Fe$]=-0.1$ to 0.1) at the average LMC metallicity of $[$Fe/H$]=-0.5$ \citep[e.g.,][]{lapenna12, vanderswaelmen13}.  Similarly, in SMC red giants $[\alpha$/Fe$]=0.0$--0.1 \citep{mucciarelli14}.  Therefore $\alpha$-elements appear to be good tracers of $[$Fe/H$]$ in these galaxies.

Of the $\alpha$-species, oxygen is the most widely-used metallicity tracer in PNe since its abundance is usually the most accurately determined.  However, models of AGB nucleosynthesis predict that at low metallicities TDU can enrich O, although the amount of enrichment differs among different AGB evolutionary codes \citep[e.g.,][]{cristallo15, ventura15b, karakas16}.  Furthermore, these calculations all show that for initial masses $\gtrsim 4$~M$_{\odot}$, O can be depleted by the CNO cycle during hot bottom burning (HBB: H-burning at the base of the convective envelope).

\citetalias{ld06} found evidence for both O enrichment and destruction in a sample of 183 LMC and SMC PNe.  We find similar effects in our abundance analysis.  In several of our observed PNe, $[$O/(S, Ar)$]=0.2$--0.4, indicating that TDU may have enhanced O in their progenitor stars.  In contrast the Type~I PN LMC~SMP~47, which likely experienced HBB based on its large N/O ratio \citepalias{ld06}, has subsolar $[$O/(S, Ar)$]$, indicating that O depletion may have occurred.  For some PNe, $[$O/(S, Ar)$]$ is solar within the abundance uncertainties, but for uniformity we use S and Ar as tracers of $[$Fe/H$]$ for all of our targets.

\subsection{Neutron-Capture Element Abundances}

In assessing whether our targets are self-enriched by \emph{s}-process nucleosynthesis, it should be noted that the initial abundances of \emph{n}-capture elements in the progenitor stars of our sample may not follow the solar abundance pattern, due to the different star formation histories and chemical evolution of the Magellanic Clouds compared to the Milky Way.  Various studies find different ratios of trans-iron element abundances relative to Fe in pre-AGB LMC stars.  The case of the SMC appears simpler, if only because there are few \emph{n}-capture element abundance determinations in its red giant stars.

Se, Kr, and Rb lie on or below the first (``light-\emph{s},'' or ls) \emph{s}-process peak, but these elements have not been detected in the spectra of late-type stars in the LMC or SMC.  Instead we compare our results with Y and Zr, two light-\emph{s} elements that are relatively well-studied in stars.  \citet{pompeia08} found $[$ls/Fe$]=-0.4$ to --0.5~dex in inner-disk LMC stars, while $[$ls/Fe$]$ is approximately solar in clusters \citep{colucci12} and the disk stars investigated by \citet{vanderswaelmen13}.  For elements belonging to the second (``heavy-\emph{s},'' or hs) peak (e.g., La), abundances from the above studies give $[$hs/Fe$]=0.2$--0.5.  In SMC Cepheids, \citet{luck98} found that $[$ls/Fe$]$ is slightly subsolar (--0.20 to --0.05 dex) while $[$hs/Fe$]$ is 0.1--0.3~dex.

Based on this information, we consider Se, Kr, and Rb to be enriched in LMC and SMC PNe if their abundances relative to S or Ar are larger than solar.  Kr is strongly enriched in all PNe in which it was detected, by 0.6~dex (SMP~63) to as much as 1.3~dex (SMP~85 and SMP~99) in LMC PNe, and by 0.6~dex in SMC~SMP~17.  Se is also enriched in five of the targets, by 0.5--0.9~dex in the LMC and 0.5~dex in SMP~17.  

The derived Se and Kr enrichment factors in these PNe generally agree with model predictions for these metallicities \citep{cristallo15, karakas16}.  Given that these models predict that elements in the second \emph{s}-process peak should be more strongly enriched at low metallicities than those in the first peak, the large Se and Kr enrichments indicate that even greater enhancements of heavier \emph{n}-capture elements can be expected.  While our nominal upper limits on [Cd/(S, Ar)] do not conform to this expectation, we note that our Cd abundances in two Galactic PNe are lower than predicted by models \citep{sterling16}, which suggests that the disagreement is likely due to uncertainties in the atomic data and/or ICF for this element.

The Type~I PN LMC~SMP~47 has both C/O and N/O ratios exceeding unity, suggesting that it experienced TDU after the cessation of HBB and has a progenitor mass $\sim$3.5~M$_{\odot} \leq M < 6$~M$_{\odot}$ \citep{ventura15b}.  The Rb abundance can also place limits on the progenitor mass, since this element can be strongly enriched if $^{22}$Ne($\alpha$, \textit{n})$^{25}$Mg dominates neutron production \citep{gh09, karakas12} as opposed to $^{13}$C($\alpha$, \textit{n})$^{16}$O, the neutron source in less massive AGB stars.  SMP~47 shows no significant enrichment of Rb, indicating that the $^{13}$C source dominated neutron production in its progenitor and that its initial mass is $\lesssim$~5.5~M$_{\odot}$ \citep{karakas16}.  This limit agrees with the predictions of \citet{ventura15b}.

The fact that the distances to the Magellanic Clouds are well known enables us to study the enrichment patterns as a function of PN luminosity and progenitor mass.  Since this is a brightness-limited sample, we preferentially selected objects at the bright end of the PNLF, whose formation mechanism has been widely debated \citep[e.g.,][and references therein]{ciardullo05}.  The substantial Se and especially Kr enrichments in the LMC PNe\footnote{Due to the detection of Se and Kr in just one SMC PN, a similar conclusion cannot be drawn at present for the SMC.} are consistent with the interpretation (based on single-star evolution) that such bright PNe are primarily produced by stars with initial masses of 2--3~M$_{\odot}$ \citep[which are expected to have the largest \emph{s}-process enrichments; e.g.,][]{cristallo15, karakas16}.  The estimated progenitor mass (1.5--2.0~M$_{\odot}$) for SMP~63 \citep{villaver03} approximately agrees with this interpretation, but progenitor masses for our other LMC targets are unknown.  Interestingly, this result appears to be at odds with the statistical analysis of \citet{badenes15}, who found that the most luminous PNe in the LMC ($L_{5007} \geq 4\times10^{34}$~erg~s$^{-1}$, corresponding to $m_{5007} \leq 18.44$) predominantly arise from stars with initial masses 1.0--1.2~M$_{\odot}$.  Binary star formation mechanisms cannot be dismissed for luminous PNe \citep[e.g.,][]{ciardullo05}, but comparisons of CNO abundances to nucleosynthesis models \citep{ventura15b} and the progenitor masses computed by \citet{villaver03, villaver07} provide evidence for a range of initial stellar masses (from $\sim$1~M$_{\odot}$ to as high as 6-8~M$_{\odot}$) for LMC PNe within a few magnitudes of the bright cutoff.  Our current sample is too small to support strong conclusions regarding progenitor mass distributions.  Determinations of \emph{s}-process enhancements in a larger number of Magellanic Cloud PNe are needed for statistically meaningful constraints on the progenitors of luminous PNe in gas-rich galaxies.

We find no significant correlations between \emph{s}-process enrichments and other nebular and stellar parameters, including C/O and N/O, central star temperature, and progenitor mass.  The small size of our sample clearly limits our ability to test these relations, but the success of our observations demonstrates that it is feasible to expand this study to other PNe in the Magellanic Clouds.  The \emph{James Webb Space Telescope} will enable such investigations to be extended to more distant Local Group galaxies, as well as to PNe well below the bright cutoff of the PNLF in these systems.  

\acknowledgments

We thank the anonymous referee, whose helpful suggestions improved this paper.  We are grateful to M.\ Reiter and R.\ Simcoe for their valuable advice regarding the reduction of FIRE data, and to J.\ Wood for her role in obtaining the Gemini data.  NCS acknowledges support from the NSF through award AST-0901432, and IUR acknowledges partial support from NSF grant PHY~14-30152 (Physics Frontier Center / JINA-CEE).  The Gemini Observatory is operated by the Association of Universities for Research in Astronomy, Inc., under a cooperative agreement with the NSF on behalf of the Gemini partnership: the National Science Foundation (United States), the National Research Council (Canada), CONICYT (Chile), Ministerio de Ciencia, Tecnolog\'ia e Innovaci\'on Productiva (Argentina), and Minist\'erio da Ci\^encia, Techologia e Inova\c{c}\~ao (Brazil).  This work has made use of NASA's Astrophysics Data System and the FRUITY Database of nucleosynthetic yields from AGB stars (fruity.oa-teramo.inaf.it).

\bibliographystyle{apj}


\begin{thebibliography}{50}
\expandafter\ifx\csname natexlab\endcsname\relax\def\natexlab#1{#1}\fi

\bibitem[{{Aller} {et~al.}(1987){Aller}, {Keyes}, {Maran}, {et~al.}}]{aller87}
{Aller}, L.~H., {Keyes}, C.~D., {Maran}, {et~al.} 1987, \apj, 320, 159

\bibitem[{{Asplund} {et~al.}(2009){Asplund}, {Grevesse}, {Sauval}, \&
  {Scott}}]{asplund09}
{Asplund}, M., {Grevesse}, N., {Sauval}, A.~J., \& {Scott}, P. 2009, \araa, 47,
  481

\bibitem[{{Badenes} {et~al.}(2015){Badenes}, {Maoz}, \&
  {Ciardullo}}]{badenes15}
{Badenes}, C., {Maoz}, D., \& {Ciardullo}, R. 2015, \apjl, 804, L25

\bibitem[{{Bernard-Salas} {et~al.}(2008){Bernard-Salas}, {Pottasch},
  {Gutenkunst}, {Morris}, \& {Houck}}]{bernard-salas08}
{Bernard-Salas}, J., {Pottasch}, S.~R., {Gutenkunst}, S., {Morris}, P.~W., \&
  {Houck}, J.~R. 2008, \apj, 672, 274

\bibitem[{{Black} \& {van Dishoeck}(1987)}]{bvd87}
{Black}, J.~H. \& {van Dishoeck}, E.~F. 1987, \apj, 322, 412

\bibitem[{{Ciardullo} {et~al.}(2005){Ciardullo}, {Sigurdsson}, {Feldmeier}, \&
  {Jacoby}}]{ciardullo05}
{Ciardullo}, R., {Sigurdsson}, S., {Feldmeier}, J.~J., \& {Jacoby}, G.~H. 2005,
  \apj, 629, 499

\bibitem[{{Colucci} {et~al.}(2012){Colucci}, {Bernstein}, {Cameron}, \&
  {McWilliam}}]{colucci12}
{Colucci}, J.~E., {Bernstein}, R.~A., {Cameron}, S.~A., \& {McWilliam}, A.
  2012, \apj, 746, 29

\bibitem[{{Cristallo} {et~al.}(2015){Cristallo}, {Straniero}, {Piersanti}, \&
  {Gobrecht}}]{cristallo15}
{Cristallo}, S., {Straniero}, O., {Piersanti}, L., \& {Gobrecht}, D. 2015,
  \apjs, 219, 40

\bibitem[{{Delgado-Inglada} {et~al.}(2014){Delgado-Inglada}, {Morisset}, \&
  {Stasi{\'n}ska}}]{delgado-inglada14}
{Delgado-Inglada}, G., {Morisset}, C., \& {Stasi{\'n}ska}, G. 2014, \mnras,
  440, 536

\bibitem[{{Delgado-Inglada} \& {Rodr{\'{\i}}guez}(2014)}]{delgado-inglada14a}
{Delgado-Inglada}, G. \& {Rodr{\'{\i}}guez}, M. 2014, \apj, 784, 173

\bibitem[{Dinerstein(2001)}]{dinerstein01}
Dinerstein, H.~L. 2001, {\apj}, 550, L223

\bibitem[{{Dopita} \& {Meatheringham}(1991)}]{dopita91b}
{Dopita}, M.~A. \& {Meatheringham}, S.~J. 1991, \apj, 377, 480 (MD91)

\bibitem[{{Dopita} {et~al.}(1994){Dopita}, {Vassiliadis}, {Meatheringham},
  {et~al.}}]{dopita94}
{Dopita}, M.~A., {Vassiliadis}, E., {Meatheringham}, S.~J., {et~al.} 1994,
  \apj, 426, 150

\bibitem[{{Dra{\v s}kovi{\'c}} {et~al.}(2015){Dra{\v s}kovi{\'c}}, {Parker},
  {Reid}, \& {Stupar}}]{dravskovic15}
{Dra{\v s}kovi{\'c}}, D., {Parker}, Q.~A., {Reid}, W.~A., \& {Stupar}, M. 2015,
  \mnras, 452, 1402

\bibitem[{{Frew} {et~al.}(2016){Frew}, {Parker}, \& {Boji{\v
  c}i{\'c}}}]{frew16}
{Frew}, D.~J., {Parker}, Q.~A., \& {Boji{\v c}i{\'c}}, I.~S. 2016, \mnras, 455,
  1459

\bibitem[{{Garc\'{i}a-Hern\'{a}ndez} {et~al.}(2009){Garc\'{i}a-Hern\'{a}ndez},
  {Iglesias-Groth}, {Rebolo}, {et~al.}}]{gh09}
{Garc\'{i}a-Hern\'{a}ndez}, J.~I., {Iglesias-Groth}, S., {Rebolo}, R., {et~al.}
  2009, \apj, 706, 866

\bibitem[{{Garc{\'{i}}a-Rojas} {et~al.}(2015){Garc{\'{i}}a-Rojas}, {Madonna},
  {Luridiana}, {et~al.}}]{garcia-rojas15}
{Garc{\'{i}}a-Rojas}, J., {Madonna}, S., {Luridiana}, V., {et~al.} 2015,
  \mnras, 452, 2606

\bibitem[{{Jacoby}(1989)}]{jacoby89}
{Jacoby}, G.~H. 1989, \apj, 339, 39

\bibitem[{{Karakas} {et~al.}(2012){Karakas}, {Garc{\'{\i}}a-Hern{\'a}ndez}, \&
  {Lugaro}}]{karakas12}
{Karakas}, A.~I., {Garc{\'{\i}}a-Hern{\'a}ndez}, D.~A., \& {Lugaro}, M. 2012,
  \apj, 751, 8

\bibitem[{{Karakas} \& {Lattanzio}(2014)}]{karakas14}
{Karakas}, A.~I. \& {Lattanzio}, J.~C. 2014, \pasa, 31, 30

\bibitem[{{Karakas} \& {Lugaro}(2016)}]{karakas16}
{Karakas}, A.~I. \& {Lugaro}, M. 2016, \apj, 825, 26

\bibitem[{Karakas {et~al.}(2009)Karakas, van Raai, Lugaro, Sterling, \&
  Dinerstein}]{karakas09}
Karakas, A.~I., van Raai, M.~A., Lugaro, M., Sterling, N.~C., \& Dinerstein,
  H.~L. 2009, {\apj}, 690, 1130

\bibitem[{{Keller} \& {Wood}(2006)}]{keller06}
{Keller}, S.~C. \& {Wood}, P.~R. 2006, \apj, 642, 834

\bibitem[{{Lapenna} {et~al.}(2012){Lapenna}, {Mucciarelli}, {Origlia}, \&
  {Ferraro}}]{lapenna12}
{Lapenna}, E., {Mucciarelli}, A., {Origlia}, L., \& {Ferraro}, F.~R. 2012,
  \apj, 761, 33

\bibitem[{{Leisy} \& {Dennefeld}(2006)}]{ld06}
{Leisy}, P. \& {Dennefeld}, M. 2006, \aap, 456, 451 (LD06)

\bibitem[{{Luck} {et~al.}(1998){Luck}, {Moffett}, {Barnes}, \&
  {Gieren}}]{luck98}
{Luck}, R.~E., {Moffett}, T.~J., {Barnes}, III, T.~G., \& {Gieren}, W.~P. 1998,
  \aj, 115, 605

\bibitem[{{Luridiana} {et~al.}(2015){Luridiana}, {Morisset}, \&
  {Shaw}}]{luridiana15}
{Luridiana}, V., {Morisset}, C., \& {Shaw}, R.~A. 2015, \aap, 573, 42

\bibitem[{{Meatheringham} \& {Dopita}(1991)}]{meatheringham91b}
{Meatheringham}, S.~J. \& {Dopita}, M.~A. 1991, \apjs, 76, 1085

\bibitem[{{Mucciarelli}(2014)}]{mucciarelli14}
{Mucciarelli}, A. 2014, Astronomische Nachrichten, 335, 79

\bibitem[{Otsuka {et~al.}(2011)Otsuka, Meixner, Riebel, {et~al.}}]{otsuka11}
Otsuka, M., Meixner, M., Riebel, D., {et~al.} 2011, {\apj}, 729, 39

\bibitem[{P\'{e}quignot \& Baluteau(1994)}]{pb94}
P\'{e}quignot, D. \& Baluteau, J.~P. 1994, {\aap}, 283, 593

\bibitem[{{Pomp{\'e}ia} {et~al.}(2008){Pomp{\'e}ia}, {Hill}, {Spite},
  {et~al.}}]{pompeia08}
{Pomp{\'e}ia}, L., {Hill}, V., {Spite}, M., {et~al.} 2008, \aap, 480, 379

\bibitem[{{Reid}(2014)}]{reid14}
{Reid}, W.~A. 2014, \mnras, 438, 2642

\bibitem[{Sharpee {et~al.}(2007)Sharpee, Zhang, Williams, {et~al.}}]{sharpee07}
Sharpee, B., Zhang, Y., Williams, R., {et~al.} 2007, {\apj}, 659, 1265

\bibitem[{{Shaw} {et~al.}(2010){Shaw}, {Lee}, {Stanghellini},
  {et~al.}}]{shaw10}
{Shaw}, R.~A., {Lee}, T.-H., {Stanghellini}, L., {et~al.} 2010, \apj, 717, 562 (S10)

\bibitem[{{Shaw} {et~al.}(2006){Shaw}, {Stanghellini}, {Villaver}, \&
  {Mutchler}}]{shaw06}
{Shaw}, R.~A., {Stanghellini}, L., {Villaver}, E., \& {Mutchler}, M. 2006,
  \apjs, 167, 201

\bibitem[{{Shortridge}(1993)}]{shortridge93}
{Shortridge}, K. 1993, in Astronomical Data Analysis Software and Systems II,
  ASP Conf.\ Ser.\ 52, ed. R.~J. {Hanisch}, R.~J.~V. {Brissenden}, \& 
  J.~{Barnes}, 219

\bibitem[{{Simcoe} {et~al.}(2013){Simcoe}, {Burgasser}, {Schechter},
  {et~al.}}]{simcoe13}
{Simcoe}, R.~A., {Burgasser}, A.~J., {Schechter}, P.~L., {et~al.} 2013, \pasp,
  125, 270

\bibitem[{{Stanghellini} {et~al.}(2009){Stanghellini}, {Lee}, {Shaw}, {Balick},
  \& {Villaver}}]{stanghellini09}
{Stanghellini}, L., {Lee}, T.-H., {Shaw}, R.~A., {Balick}, B., \& {Villaver},
  E. 2009, \apj, 702, 733

\bibitem[{{Stanghellini} {et~al.}(2003){Stanghellini}, {Shaw}, {Balick},
  {Mutchler}, {Blades}, \& {Villaver}}]{stanghellini03}
{Stanghellini}, L., {Shaw}, R.~A., {Balick}, B., {Mutchler}, M., {Blades},
  J.~C., \& {Villaver}, E. 2003, \apj, 596, 997

\bibitem[{Sterling \& Dinerstein(2008)}]{sterling08}
Sterling, N.~C. \& Dinerstein, H.~L. 2008, {\apjs}, 174, 158

\bibitem[{{Sterling} {et~al.}(2016){Sterling}, {Dinerstein}, {Kaplan}, \&
  {Bautista}}]{sterling16}
{Sterling}, N.~C., {Dinerstein}, H.~L., {Kaplan}, K.~F., \& {Bautista}, M.~A.
  2016, \apjl, 819, L9

\bibitem[{{Sterling} {et~al.}(2015){Sterling}, {Porter}, \&
  {Dinerstein}}]{sterling15}
{Sterling}, N.~C., {Porter}, R.~L., \& {Dinerstein}, H.~L. 2015, \apjs, 218, 25

\bibitem[{{Tsamis} {et~al.}(2003){Tsamis}, {Barlow}, {Liu}, {Danziger}, \&
  {Storey}}]{tsamis03}
{Tsamis}, Y.~G., {Barlow}, M.~J., {Liu}, X.-W., {Danziger}, I.~J., \& {Storey},
  P.~J. 2003, \mnras, 345, 186 (T03)

\bibitem[{{Van der Swaelmen} {et~al.}(2013){Van der Swaelmen}, {Hill},
  {Primas}, \& {Cole}}]{vanderswaelmen13}
{Van der Swaelmen}, M., {Hill}, V., {Primas}, F., \& {Cole}, A.~A. 2013, \aap,
  560, 44

\bibitem[{{Ventura} {et~al.}(2015){Ventura}, {Stanghellini}, {Dell'Agli},
  {Garc{\'{\i}}a-Hern{\'a}ndez}, \& {Di Criscienzo}}]{ventura15b}
{Ventura}, P., {Stanghellini}, L., {Dell'Agli}, F.,
  {Garc{\'{\i}}a-Hern{\'a}ndez}, D.~A., \& {Di Criscienzo}, M. 2015, \mnras,
  452, 3679

\bibitem[{{Villaver} {et~al.}(2003){Villaver}, {Stanghellini}, \&
  {Shaw}}]{villaver03}
{Villaver}, E., {Stanghellini}, L., \& {Shaw}, R.~A. 2003, \apj, 597, 298

\bibitem[{{Villaver} {et~al.}(2004){Villaver}, {Stanghellini}, \&
  {Shaw}}]{villaver04}
---. 2004, \apj, 614, 716

\bibitem[{{Villaver} {et~al.}(2007){Villaver}, {Stanghellini}, \&
  {Shaw}}]{villaver07}
---. 2007, \apj, 656, 831

\bibitem[{{Wood} {et~al.}(2006){Wood}, {Dinerstein}, {Geballe}, \&
  {Sterling}}]{wood06}
{Wood}, J.~L., {Dinerstein}, H.~L., {Geballe}, T.~R., \& {Sterling}, N.~C.
  2006, BAAS, 38, 1113

\end{thebibliography}

\end{document}